\documentclass[manuscript]{aastex}
\shorttitle{Particle anisotropies at the boundary}
\shortauthors{Florinski}

\begin{document}

\title{Entering the ``magnetic highway'': energetic particle anisotropies at
the heliospheric boundary}

\author{V. Florinski}

\affil{Department of Physics and Center for Space Plasma and Aeronomic Research,
University of Alabama, Huntsville, AL 35899}

%-------------------------------------------------------------------------------
\begin{abstract}
In August of 2012 the Voyager 1 space probe entered a distinctly new region of
space characterized by a virtual absence of heliospheric energetic charged
particles and magnetic fluctuations, dubbed a ``magnetic highway''.
Prior to their disappearance, the particle distributions strongly peaked at a
90$^\circ$ pitch angle implying a faster particle escape along the magnetic
field lines.
We investigate the process of particle crossing from the heliosheath region into
the ``magnetic highway'' region using a kinetic approach resolving scales of the
particle's cyclotron radius and smaller.
We show that a ``loss-cone'' type distribution naturally arises as the orbiting
particles enter a region of space with an extremely low pitch-angle scattering
rate.
\end{abstract}


%-------------------------------------------------------------------------------
\section{Introduction}
Since the end of 2004, the Voyager 1 space probe has been exploring the region
of space downstream of the solar wind termination shock, known as the
heliosheath \citep[e.g.,][]{stone05, decker05}.
Observations showed that the heliosheath plasma is generally in a turbulent
state \citep{burlaga10} with the mean magnetic field magnitude increasing with
heliocentric distance from some 1 $\mu$G just beyond the shock (94 AU) to about
3 $\mu$G by mid-2012 when the spacecraft was at 122 AU from the Sun
\citep{burlaga12}.
The heliosheath is permeated by energetic ions with energies below several MeV,
produced inside the heliosphere from interstellar material (neutral atoms) and
accelerated at the termination shock \citep[e.g.,][]{decker08, florinski09a}.

Starting in August of 2012, the HEP intensity experienced a series of rapid
drops and increases after which they disappeared altogether \citep{stone12,
decker12b}.
The last event was accompanied by a jump in the magnetic field strength from
3 $\mu$G to 4 $\mu$G without a change in direction \citep{burlaga12}.
Remarkably, the magnetic field became very smooth after crossing the boundary
with turbulent fluctuations virtually absent.
Such a change in the properties of the magnetic field was expected to happen
across the heliopause since it was generally thought that interstellar
turbulence existed on vastly greater scales \citep{armstrong95}.
The absence of magnetic field rotation casts some doubt on the interpretation
that the boundary just crossed is indeed the heliopause.
It is clear, however, that particle escape along the field lines in the new
region is very efficient, which prompted the name ``magnetic highway'' (MH).
It is not clear at this time whether the new region is part of the heliosphere
or the interstellar cloud around it, because of the nonfunctional plasma
instrument on Voyager 1.
In the scenario discussed below this difference is of no significance as long as
the magnetic field remains turbulence free beyond the boundary.

In this Letter we simulate the process of energetic particle crossing from a
turbulent region (the heliosheath) into a laminar region (the MH or LISM).
Specifically, we propose an explanation for the observational result that the
intensity of particles streaming along the magnetic field decreased faster than
those gyrating at nearly $90^\circ$ angles resulting in a double loss cone
pitch-angle distribution \citep{cummings12}.
Because the HEP intensity changes occurred on short time scales (i.e.,
comparable to $V_{V1}/r_g$, where $V_{V1}$ is Voyager 1 speed and $r_g$ is the
cyclotron radius of a typical HEP ion), a model must be capable of resolving
scales of $r_g$ and smaller.
A diffusive approach is obviously invalid here because the particle populations
were highly anisotropic immediately beyond the boundary.
Using a fully kinetic approach in a simple numerical model we show that a
flattened pitch-angle distribution is a natural consequence of gyrating
particles traveling farther from the boundary before escaping under scatter-free
conditions.

%-------------------------------------------------------------------------------
\section{The kinetic transport model}
\begin{figure}
\plotone{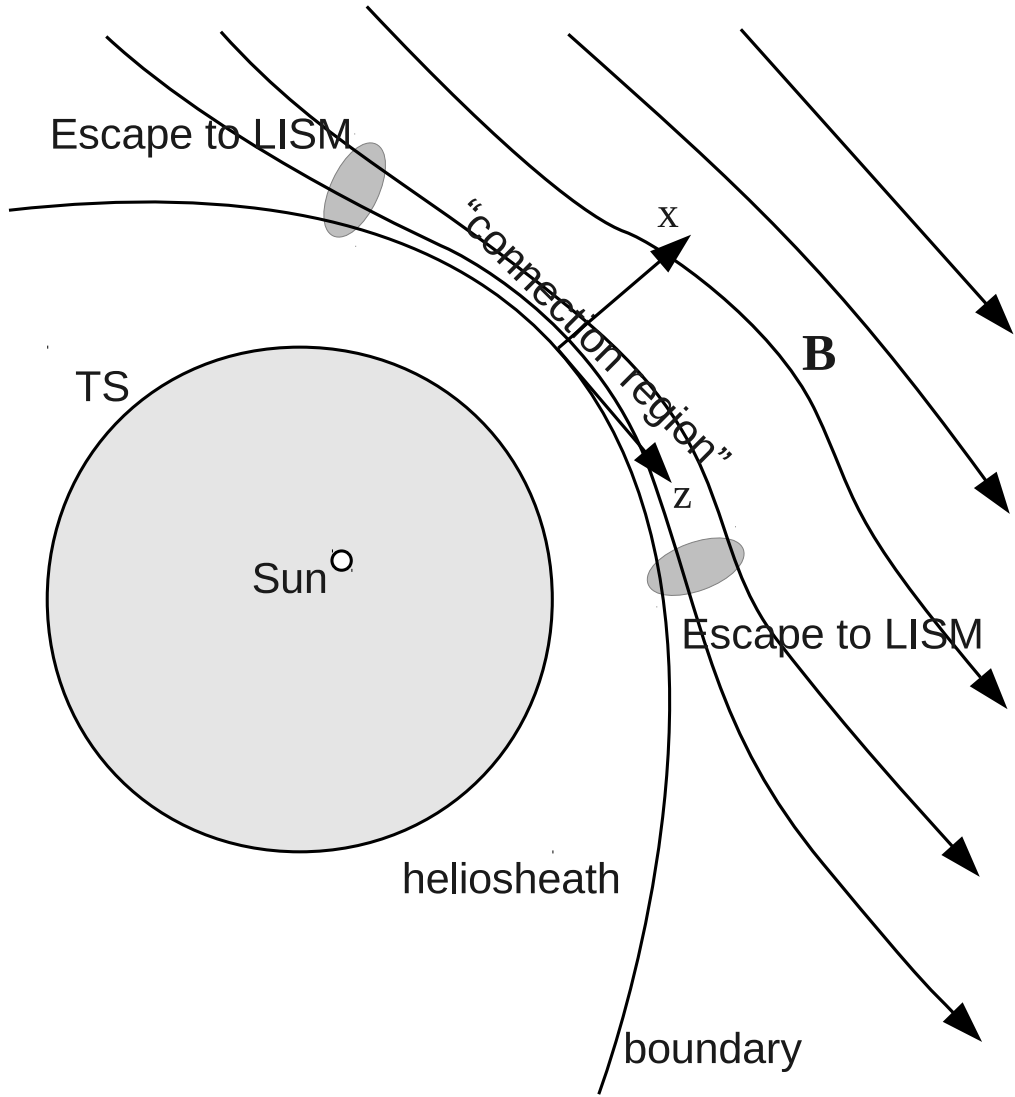}
\caption{An illustration of the particle transport model geometry.
We consider a narrow ``connection region'' where the magnetic field presses
against the boundary.
\label{fig_draping}}
\end{figure}

The large-scale geometry of our model is illustrated schematically in Figure
\ref{fig_draping}.
We consider a narrow region immediately adjacent to the boundary on either side.
The $x$-axis is normal to the boundary and the $z$-axis is in the direction of
the mean magnetic field.
It is assumed that the magnetic field lines in the MH region are pressed against
the boundary for a distance $z_\mathrm{max}$ (the ``connection region'' in
Fig. \ref{fig_draping}) after which they separate from the boundary.
A particle reaching the end of the connection region is assumed to escape the
region and is removed from the system.

The background plasma velocity indirectly measured by Voyager 1 was very small
during 2010--2012 \citep{krimigis11, decker12a}.
For simplicity we assume here that the plasma velocity is zero on both sides of
the boundary (this assumption is not critical as long as the plasma speed is
much less than the particle velocity $v$, which is usually satisfied everywhere
in the heliosheath).
We take the mean field to be $B_\mathrm{HS}=3$ $\mu G$ and $B_\mathrm{MH}=4$
$\mu G$ on the heliosheath side, and the MH side of the boundary, respectively.
The general Boltzmann equation describing energetic particle transport in a
magnetized plasma with turbulent fluctuations may be written as
\begin{equation}
\frac{\partial f}{\partial t}
+v\sqrt{1-\mu^2}\cos\varphi\frac{\partial f}{\partial x}
+v\mu\frac{\partial f}{\partial z}
-\Omega\frac{\partial f}{\partial\varphi}=Sf,
\end{equation}
where $f(x,z,\mu,\varphi)$ is the phase space density, $\mu$ is the pitch-angle
cosine, $\varphi$ is the gyrophase, and $\Omega$ is the cyclotron frequency.
The operator $S$, describing particle scattering in solid angle, could be
obtained from the theory of wave-particle interaction \citep[e.g.,][]{jokipii72}.
The transport across the magnetic field is produced entirely by scattering of
the particles in gyrophase.
Because the plasma is stationary, there is no electric field and the energy of
the particles is assumed to be conserved.

For simplicity we adopt a model with an isotropic scattering operator,
\begin{equation}
Sf=D\frac{\partial}{\partial\mu}\left[(1-\mu^2)\frac{\partial f}{\partial\mu}\right]
+\frac{D}{1-\mu^2}\frac{\partial^2f}{\partial\varphi^2},
\end{equation}
where $D$ is the scattering coefficient.
The parallel mean free path corresponding to (2) is $\lambda_\parallel=v\tau$,
where $\tau=(2D)^{-1}$ is the scattering rate.
For large $\Omega\tau$ used here the transport is predominantly along the
magnetic field \citep[e.g.,][]{forman74}.
Equation (1) is solved numerically along the characteristics using a
stochastic integration method \citep[e.g.,][]{mackinnon91, florinski09b}.
Instead of using Eq. (2) directly, we use the random walk on a sphere technique
of \citet{ellison90} that avoids the singularity at $\mu=\pm 1$.
For small scattering angles the process is identical to a two-dimensional random
walk on a plane.

%-------------------------------------------------------------------------------
\section{Intensity and pitch-angle distributions}
\begin{figure}
\plotone{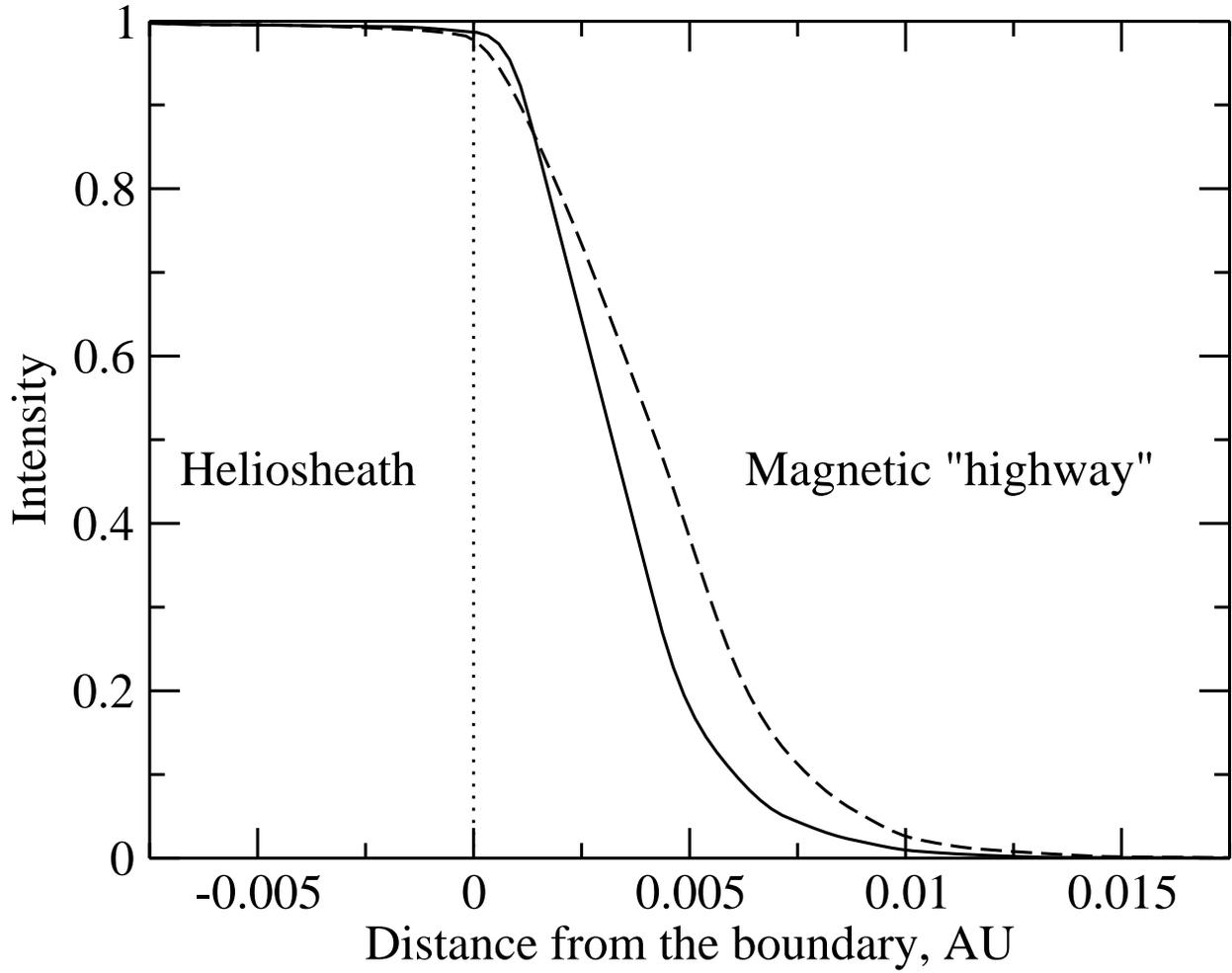}
\caption{Intensity profiles of $\mu=1$ (solid line) and $\mu=0$ (dashed line)
particles as a function of distance from the boundary (the dotted line).
\label{fig_hpcross}}
\end{figure}

We performed a series of simulations varying the scattering rates,
$D_\mathrm{HS}$ and $D_\mathrm{MH}$, and the length of the connection regions
$z_\mathrm{max}$, within a sufficiently wide range.
Figure \ref{fig_hpcross} shows a typical result for 1 MeV protons obtained
with $D_\mathrm{HS}=10^{-3}$ s$^{-1}$, $D_\mathrm{MH}=2\times 10^{-6}$
s$^{-1}$, and $z_\mathrm{max}=32$ AU.
Such a large difference in scattering rates is required to produce a noticeable
difference between particles traveling at different pitch angles.
The MH region is essentially scatter-free ($\lambda_\parallel=23$ AU) and
particles are only able to penetrate a distance comparable to $r_g=0.0024$ AU
before escaping into the LISM.
Intensity profiles of $\mu=1$ and $\mu=0$ ions are shown with a solid and dashed
lines, respectively.

The particle distribution is assumed to be isotropic at the left (heliosheath)
boundary of the simulation and the two curves are almost identical up to $x=0$.
Afterwards, ions streaming along the magnetic field diminish earlier than ions
near $90^\circ$ pitch angle.
The right (MH) boundary plays no role in the simulation because all particles
exit the simulation box before reaching it.
The difference in penetration depth is of the order of a cyclotron radius in the
MH.
We note that the result remains essentially unchanged if the ratio
$D_\mathrm{MH}/z_\mathrm{max}$ is kept a constant.
This means that the model applies for larger or smaller scattering rates as long
as the unknown escape distance is adjusted accordingly.

\begin{figure}
\plotone{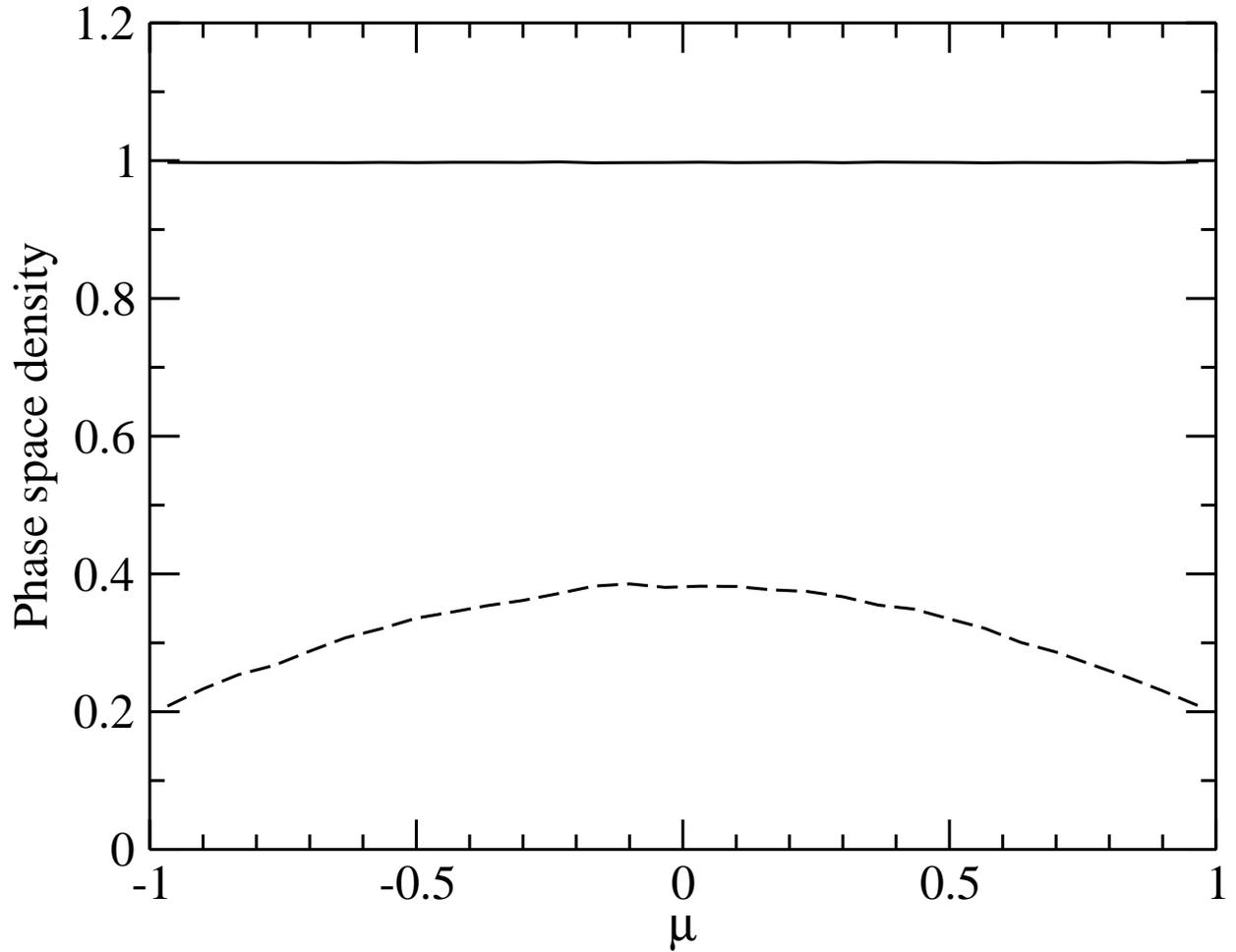}
\caption{Pitch-angle distributions of HEPs in the heliosheath ($x=-0.005$ AU,
solid line) and in the MH region ($x=0.005$ AU, dashed line)
The intensity is normalized to one at the left boundary of the simulation.
\label{fig_pitch}}
\end{figure}

Figure \ref{fig_pitch} plots the HEP pitch-angle distributions on the
heliosheath side at $x=-0.005$ AU (solid line) and on the MH side at $x=0.005$
AU (dashed line).
The initially isotropic ion population becomes highly anisotropic past the
boundary as it becomes depleted in $\mu=\pm 1$ particles.
Some $\mu=0$ particles have also scattered and escaped from the system, so that
the total intensity is significantly reduced.
This result is consistent with Voyager 1 observations showing a flattened
(pancake like) pitch-angle ion distribution beyond the boundary
\citep{cummings12, decker12b}.

%-------------------------------------------------------------------------------
\section{Discussion}
In this Letter we proposed a simple model to explain the double loss cone type
distributions of MeV heliosheath particles observed immediately beyond the
boundary crossed by Voyager 1 in 2012.
The model is based on a concept that particle scattering rates are much smaller
in the MH region than in the heliosheath.
Particles gyrating near $\mu=1$ are able to penetrate deeper into the new region
because of their larger gyro-radius, whereas particles streaming along the field
rapidly escape into the LISM.
The model does not require magnetic mirrors on either side of the Voyager
crossing point, but only that there exist a boundary magnetic connection region
of considerable extent (a few tens of AU).
Our model is unrelated to the concept proposed by \citet{mccomas12}, where the
drop in energetic particle flux is a consequence of Voyager 1 leaving the region
magnetically connected to the termination shock.

Our very simple model qualitatively reproduces the observed changes in particle
distribution across the boundary.
The same model may be applied to ions with higher and lower energies by
adjusting the scattering rate accordingly.
The contrast between particles traveling at 0 and $90^\circ$ pitch angles in our
model appears to be less than observed.
This is primarily due to a strictly perpendicular magnetic field geometry used
here.
The model therefore lacks diffusive transport across the magnetic field from
magnetic field line meandering.
A proper statistical treatment of this effect on kinetic scales requires a
prescription of the complete three-dimensional turbulent magnetic field, which
is beyond the scope of this paper.
Magnetic field line diffusion would make particle intensity decrease at a slower
rate.
Consequently, the HEPs may be still around several $r_g$ deep into the MH
region.

Another simplification is our use of a very thin boundary between the two
regions.
The boundary could be diffuse and/or turbulent, which would also make the HEP
intensity decrease slower than predicted here.
It should be pointed out that the true radial extent of the particle decrease
region is unknown because of a large uncertainty in the background plasma speed
measurement.
If the flow was slightly anti-sunward, the region could be very thin (perhaps of
the order of $r_g$).
Irrespective to this result, we think that the basic premise of our model
(particle escape into the LISM at different rates) adequately describes the
fundamental physics of the newly discovered boundary.

\acknowledgments
This work was supported, in part, by NASA grants NNX10AE46G, NNX12AH44G, NSF
grant AGS-0955700 and by a cooperative agreement with NASA Marshall Space Flight
Center.

%-------------------------------------------------------------------------------

\end{document}